\documentclass[aps,nofootinbib,twocolumn]{revtex4}

\usepackage{amssymb,amstext,amsmath,amsthm}
\usepackage[dvips]{graphicx}
\usepackage{latexsym}
\usepackage{psfrag}
\usepackage{amsfonts}
\usepackage{bbm}

\def\nn{\nonumber}

\newcommand{\eqa}{\begin{eqnarray}}
\newcommand{\neqa}{\end{eqnarray}}
\newcommand{\ba}{\begin{eqnarray}}
\newcommand{\ea}{\end{eqnarray}}
\newcommand{\equ}{\begin{equation}}
\newcommand{\nequ}{\end{equation}}

\def\q{\quad}

\usepackage{bbm}

\begin{document}

\title{\Large\bf  Regge calculus from a new angle}

\author{Benjamin Bahr$^1$ and Bianca Dittrich$^2$\\
\small $^1$ DAMTP, University of Cambridge,\\
\small  Wilberforce Road, Cambridge CB3 0WA, UK \\
\small   $^2$ MPI f. Gravitational Physics, Albert Einstein Institute,\\
 \small Am M\"uhlenberg 1, D-14476 Potsdam, Germany\\
 }

\date{\small\today}

\begin{abstract}

In Regge calculus space time is usually approximated by a triangulation with flat simplices. We present a formulation using simplices with constant sectional curvature adjusted to the presence of a cosmological constant. As we will show such a formulation allows to replace the length variables by 3d or 4d dihedral angles as basic variables. Moreover we will introduce a first order formulation, which in contrast to using flat simplices, does not require any constraints. These considerations could be useful for the construction of quantum gravity models with a cosmological constant.

\end{abstract}

\maketitle

\section{Introduction}

Regge calculus \cite{Regge} is an elegant discrete formulation of general relativity, where space time is approximated by a piecewise flat (simplicial) manifold. Beside applications in numerical relativity it has been used in quantum gravity as a starting point for a non-perturbative definition for path integral formulations \cite{RuthRegge,Hamber}. Whereas originally Regge calculus has been based on length variables, newer developments suggest that for four--dimensional gravity other variables might be preferable, for instance in order to define path integral quantization or canonical formulations \cite{bdjr}. In particular, spin foam models are rather first order formulations and additionally use areas (and 3d angles) instead of length variables. Attempts to base Regge calculus on exclusively area variables \cite{carloarea,Makela,Barrett,Wainwright,RuthRegge} have to face the difficulty, that the constraints that ensure that a piecewise geometry can be uniquely defined, are very non--local and not known explicitly. As has been recently shown \cite{bdss} this can be circumvented by introducing additional variables, namely the 3d dihedral angles in the tetrahedra of the triangulation.

Barrett \cite{barrett} introduced a first order formulation for Regge calculus, using length and 4d dihedral angles  as variables. In order to ensure that the 4d dihedral angles are consistent with the geometry defined by the length variables one constraint per 4--simplex had to be added to the action. By combining \cite{bdss} and \cite{barrett} one can obtain a formulation with areas, 3d angles and 4d angles.

Another development \cite{bbbd2} concerns an improvement of the Regge action for vacuum gravity with a cosmological constant.  Usually the cosmological constant is accommodated by adding a volume term to the action and by still using a piecewise flat triangulation, i.e. simplices which have flat geometry. In this way even the simplest solution corresponding to homogeneously (maximally symmetric) curved space is only an approximate one compared to the continuum solution. The reason is that one uses flat simplices to approximate (homogeneously) curved space. Here the idea is to use simplices with homogeneous curvature, i.e. constant sectional curvature, instead, so that the discrete solution with homogeneous curvature is also an exact solution of the continuum. It has been show in \cite{bbbd2} that this allows a better representation of the (diffeomorphism) symmetries of general relativity and therefore could simplify quantization. Indeed in 3d the Tuarev--Viro model \cite{tuarev} gives a partition function for homogeneously curved simplices \cite{taylor}, whereas a similar quantization based on flat simplices and with a cosmological constant is not available yet.  In general using a (positive) cosmological constant has the advantage that it can serve as a regulator in the path integral, as for instance the translation symmetries which for flat simplices would lead to divergencies are now compactified to the (4-) sphere.

The geometry of simplices with and without (homogeneous) curvature differs in one important way: For flat simplices any set of angles can specify at most the conformal geometry of the simplex (i.e. all the lengths modulo one factor). Moreover the dihedral angles of a flat simplex have to satisfy one constraints, namely that the determinant of the so--called Gram matrix vanishes. This is the constraint that is added in \cite{barrett} to obtain a first order formulation. For homogeneously curved simplices, however, the set of dihedral angles specifies the full geometry of the simplex. Moreover they do not have to specify any constraints.\footnote{Except for certain inequalities, that can be understood to replace the generalized triangle inequalities for the length variables. For a spherical simplex the angle Gram matrix, defined in the appendix \ref{appendix}, has to be positive definite. For a hyperbolic $D$-simplex the determinant of the angle Gram matrix has to be negative, and all principle $ D\times D$ sub-matrices have to be positive definite and the cofactors positive, see for instance \cite{feng06} .} As we will see this will allow us to obtain formulations where the basic variables are only angles, that is either the 3d dihedral angles or the 4d dihedral angles.  Hence in a path integral the integration over length variables can be replaced by an integration over angles, which opens up new ways for quantization, for instance for constructing the path integral measure. In spin foam models one usually integrates out the metric variable and is just left with a connection variable, which would correspond to using only the 4d dihedral angles.

In the next section \ref{first} we will present a first order formulation. Here we will see that the formulation with curved simplices is far less complicated compared to working with flat simplices. In section \ref{3d} we will introduce a formulation based on 3d dihedral angles. These have to be constrained, and we will see that these constraints are again relations between angles, this time 2d angles. Finally we discuss in section \ref{4d} a formulation with 4d dihedral angles, which have also to be constrained, this time the constraints are relations between 3d angles. We conclude in section \ref{conc} and summarize the necessary background on geometric relations in simplices, in particular relations between dihedral angles in the appendix \ref{appendix}.

\section{A first order formulation}\label{first}

We will start to describe a first order formulation. We will see that in contrast to using flat simplices \cite{barrett} we do not need to add any constraints to the action and to obtain the equations is extremely straightforward. To use simplices with constant curvature for a first order formulation was suggested in \cite{barrett}, but neither an action nor any other details have been given.

The Regge action\footnote{We will work with Euclidean signature.} in 4 dimensions with simplices of constant curvature and cosmological constant $\Lambda=3\kappa$  (and without boundary terms) is given by \cite{bbbd2}
\eqa\label{reg1}
S[l_e]=\sum_t a_t(l) \epsilon_t (l) + 3\kappa \sum_\sigma V_\sigma(l) \q .
\neqa
Here we use lengths $l_e$ associated to edges as basic variables.
The subindex $t$ denotes triangles, $a_t$ is the area of the triangle $t$ and $\epsilon_t:=2\pi-\sum_{\sigma \supset t} \theta^\sigma_t$ is the deficit angle associated to the triangle $t$. Moreover, $\theta^\sigma_t$ is the 4d dihedral angle in the simplex $\sigma$ between the two tetrahedra that share the triangle $t$. The deficit angles specify the (corrected) curvature, which has distributional support on the triangles. The full curvature is given by these deficit angles and a homogeneous contribution. The latter leads to a plus sign in front of the volume term in the action (\ref{reg1}) as compared to the continuum expression $\int \sqrt{g} (R-2\Lambda)d^4x$ for the Einstein Hilbert action, where $R$ is the (full) Ricci curvature scalar.

The variation with respect to the edge length $l_e$ of the action (\ref{reg1}) gives the equations of motion
\eqa\label{reg2}
\sum_{t \supset e}  \frac{\partial a_t}{\partial l_e} \epsilon_t =0 \q .
\neqa
The variation of the deficit angles and the volume cancel out due to the Schl\"afli identity
\eqa\label{schlaefli}
3\kappa \delta V_\sigma =\sum_{t\subset \sigma} a_t \delta \theta^\sigma_t
\neqa
which holds for any variation $\delta$ of the geometry of the 4--simplex $\sigma$.

In a $3+1$ formulation the 4d dihedral angles $\theta^\sigma_t$ would specify the extrinsic curvature and can therefore be taken as first order variables. Indeed this has been also done in \cite{barrett} for flat simplices. As already mentioned in contrast to the dihedral angles in a flat simplex the dihedral angles in a simplex with constant curvature do not need to satisfy any constraints. Indeed it turns out that we do get the correct equations of motion if we express the volume term (and the deficit angles) in (\ref{reg1}) in terms of the dihedral angles:
\eqa\label{reg3}
S[l_e,\theta^\sigma_t]=\sum_t a_t(l) \epsilon_t (\theta) + 3\kappa \sum_\sigma V_\sigma(\theta^\sigma) \q .
\neqa
Now varying with respect to $l_e$ gives the same equation as before (\ref{reg2})
\eqa\label{reg4}
\sum_{t \supset e}  \frac{\partial a_t}{\partial l_e} \epsilon_t =0
\neqa
whereas the variation with respect to $\theta^\sigma_t$ gives
\eqa\label{reg5}
-a_t(l_e)+a_t(\theta^\sigma)=0
\neqa
where we use the Schl\"afli identity (\ref{schlaefli}) to find the variation of the volume term. Note that the equation of motion (\ref{reg5}) accomplishes two tasks at once, firstly it ensures that the dihedral angles define consistently a simplicial geometry (that is the 4--simplices glue properly together), and secondly that this geometry coincides with the one defined by the length variables.\footnote{As discussed in \cite{barrett} one has to take the possibility into account that the ten areas of a 4--simplex might allow several length assignments. This is however only a discrete ambiguity.}.

\section{A formulation with 3d dihedral angles} \label{3d}

As mentioned in the introduction spin foam models suggest to use areas and 3d dihedral angles instead of length variables as basic variables. These variables can be easily constructed out of gauge formulations for gravity, such as the Plebanski formulation \cite{plebanski} on which spin foams are based, see also \cite{bdjr}. Such a formulation was developed in \cite{bdss} for flat simplices. The basic idea is to start with the geometry of the tetrahedra. The geometry of one tetrahedron is described by six lengths or equivalently by the four areas and six 3d dihedral angles satisfying four Gau\ss~constraints. These constraints allow to express four of the six dihedral angles as a function of the four areas and the remaining two angles (which have to be non--opposite). We take the same starting point for the curved simplices. Here the difference with the flat case is that we can express the four areas as a function of the six 3d dihedral angles (see appendix \ref{details}). So one can either decide to keep the areas and to add these expressions as constraints or to just work with the six dihedral angles per tetrahedron. Following the latter route we have furthermore to ensure that the tetrahedra in one simplex properly glue together. That is the geometry of the triangle shared by two tetrahedra has to coincide if defined by the two sets of dihedral angles associated to the two tetrahedra. Again since the three angles in a triangle determine the geometry it is sufficient to ensure that these three 2d angles coincide if calculated from the two sets of 3d dihedral angles. Consider a 4--simplex and label the vertices by $p=1,\ldots,5$. As shown in the appendix the relation between 3d dihedral angles and 2d dihedral angles in a $4$-simplex $\sigma$ is
\eqa\label{ba3}
\cos\alpha_{lm,kp}(\phi)=\frac{\cos\phi_{lm,p}+\cos\phi_{kl,p}\cos\phi_{km,p}}{\sin\phi_{kl,p}\sin\phi_{km,p}}\q .
 \neqa
 Here $\phi_{lm,p}$ is the dihedral angle in a tetrahedron $\sigma(\hat p)$ opposite the vertex $p$ and at the edge opposite the edge $(lm)$. The $\alpha_{lm,kp}$ is the 2d dihedral angle in the triangles $\sigma(\hat k \hat p)$ opposite the edge $(kp)$ and at the vertex opposite the edge $(lm)$.
The constraints that have to hold between the 3d angles
$\phi$ are as follows: the angles $\alpha$ in the triangle
$\sigma(\hat k \hat p) $ have to coincide if computed from either the 3d
angles in the tetrahedron $\sigma(\hat k)$ or the 3d angles in the
tetrahedron $\sigma(\hat p)$, that is
 \ba\label{ba4}
C_{lm,kp}:=\cos\alpha_{lm,kp}(\phi_{\cdot\cdot,p})-\cos
\alpha_{lm,pk}(\phi_{\cdot\cdot,k}) \, .
\ea
As we started with the geometry of the tetrahedra and ensured that these glue properly to 4--simplices, we also enforced that the 4--simplices properly glue together: For this the geometry of every tetrahedron shared by two simplices has to coincide. This is satisfied by construction, as we took the geometry of the tetrahedra as independent variables (and constrained it afterwards to ensure gluing to 4--simplices).

Finally the action for 4d Regge calculus with curved simplices
based on 3d dihedral angles $\phi$ and with Lagrange multipliers
$\lambda^\sigma_{e,e'}$ is
\ba
 S&=&\sum_{t} a_t(\phi) \epsilon_t(\phi)
+3\kappa \sum_{\sigma} V_{\sigma}(\phi)  +\nn\\
&&\sum_\sigma \sum_
{e,e'\subset \sigma} \lambda_{e,e'}^\sigma C^{\sigma}_{e,e'} (\phi)
\ea
where $C^{\sigma}_{e,e'}$ is zero, if the edges $e,e'$ do not
share a triangle, and coincides with $C_{lm,kp}$ for $e,e'$ the edges
opposite the triangles $(lmk)$ and $(lmp)$ respectively. In the appendix \ref{details} we will give formulas for the definition of the areas and 4d dihedral angles as a function of the 3d dihedral angles. Note that a priori one has to specify how to calculate these quantities from the 3d dihedral angles, e.g. for an area $a_t$ one has to define the tetrahedron, whose set of 3d angles is used to calculate the area from. (Alternatively take the average over all adjacent tetrahedra.)  The same holds for the volumes $V_\sigma$, that is one can first specify how to calculate lengths from the 3d dihedral angles, which can then in principle used to find the volume. (Again one possibility is to take as these lengths the average of the length variables calculated from the dihedral angles of the adjacent tetrahedra in the simplex $\sigma$.) Different choices lead to the same result if the constraints are satisfied.

The constraints ensure that one can calculate consistently length variables from the 3d angles. The inverse solutions $\phi(l)$ can be used in the action to re-obtain the action (\ref{reg1}). Hence these two action lead to the same equations of motions. For the same reason if we introduce the (first order) variables $\theta^\sigma_t$ and express the deficit angles $\epsilon_t$ and the volumes $V_\sigma$ as functions of these variables, we obtain a first order formulation equivalent to (\ref{reg3}) which uses only angles as variables.

\section{A formulation with 4d dihedral angles}\label{4d}

Similarly one can obtain a formulation involving only 4d dihedral angles (and Lagrange multipliers enforcing constraints).

The ten dihedral angles of a 4--simplex determine uniquely its geometry and for a single 4--simplex we can take these 4d dihedral angles as free variables. If we glue the simplices together, we have to ensure that this can be consistently done. Two neighboring 4--simplices might induce a priori different geometries for  the common tetrahedron. We have to introduce constraints that ensure that this geometry coincides. To this end we use the formula expressing the 3d dihedral angles as a function of the 4d dihedral angles
\ba\label{4d1}
  \cos\phi_{lm,p}(\theta)&=&
\frac{\cos\theta_{lm}+\cos\theta_{pl}\cos\theta_{pm}}{\sin\theta_{pl}\sin\theta_{pm}}\, .
 \ea
Here $\phi_{lm,p}$ is the dihedral angle in a tetrahedron $\sigma(\hat p)$ opposite the vertex $p$ and at the edge opposite the edge $(lm)$. The 4d dihedral angle $\theta_{pm}$ is the one at the triangle $\sigma(\hat p \hat m)$ opposite the edge $(pm)$.

The geometry of the common tetrahedron is fixed by the six 3d dihedral angles, hence we introduce one constraint per tetrahedron and edge in this tetrahedron ensuring that these 3d dihedral angles coincide if calculated from either of the two adjacent 4--simplices $\sigma$ and $\sigma'$:
\ba\label{4d2}
C^\tau_e(\theta^\sigma,\theta^{\sigma'})=\cos \phi^\tau_e(\theta^\sigma) -\cos \phi^\tau_e(\theta^{\sigma'})  \q
\ea
where $\phi^\tau_e$ is the 3d dihedral angle at the edge $e$.
The Regge action is now
\ba\label{4d3}
 S&=&\sum_{t}   \sum_{\sigma\supset t}   a_t^\sigma(\theta^\sigma)\left( \frac{2\pi}{N_t} -  \theta^\sigma_t\right)
+3\kappa \sum_{\sigma} V_{\sigma}(\theta^\sigma)  +\nn\\
&&\sum_\tau \sum_
{e\subset \tau} \lambda_{e}^\tau C^{\tau}_{e} (\theta^\sigma,\theta^{\sigma'})  \q .
\ea

Here $N_t$ is the number of simplices adjacent to the triangle $t$ so that the first term in (\ref{4d3}) gives the average of the areas $a_t^\sigma$ attached to the same triangle but computed from the dihedral angles of the 4--simplices $\sigma$ adjacent to the triangle (see appendix \ref{details}). Again the constraints allow to determine consistently the set of edge length so that the functions $\theta^\sigma_t(l)$ can be reinserted into the action and one would obtain the same equation of motions as for the original action (\ref{reg1}). We can however also vary (\ref{4d3}) directly with respect to the 4d dihedral angles $\theta^{\sigma'}_{t'}$. Here we can again use the Schl\"afli identity and obtain
\ba\label{4d4}
\sum_{t\subset \sigma'} \frac{\partial a_t(\theta^{\sigma'})}{ \partial \theta^{\sigma'}_{t'}} \left(\frac{2\pi}{N_t}-\theta^{\sigma'}_t\right) + \sum_{\tau\subset \sigma'} \sum_{e\in \tau} \lambda^\tau_e \frac{\partial C^\tau_e}{\partial \theta^{\sigma'}_{t'}}=0 \, .\q
\ea
Now we multiply this equation (\ref{4d4}) by $\frac{\partial \theta^{\sigma'}_{t'}}{\partial l^{\sigma'}_{e'}}$, i.e. the inverse to $\frac{\partial l^{\sigma'}_{e'}}{\partial \theta^{\sigma'}_{e'}}$, where $l_{e'}^{\sigma'}$ is the edge length of $e'$ as computed from the dihedral angles in $\sigma'$, and sum over $t'\subset \sigma'$:
\ba\label{4d5}
\sum_{t\subset \sigma'} \frac{\partial a_t(\theta^{\sigma'})}{ \partial l^{\sigma'}_{e'}}\left(\frac{2\pi}{N_t}-\theta^{\sigma'}_t\right) + \sum_{\tau\subset \sigma'} \sum_{e\in \tau} \lambda^\tau_e \frac{\partial C^\tau_e}{\partial l^{\sigma'}_{e'}}=0 \, .\q
\ea

Finally summing over all $\sigma' \supset e'$ the contribution with the derivatives of the constraints cancel if the constraints are satisfied. (The sum over the simplices gives two contributions per tetrahedron and edge with opposite sign. If the constraints are satisfied these contributions cancel.) On the constraint hypersurface (where lengths and areas as computed from different simplices agree) we recover the equations of motion (\ref{reg2})
\ba
\sum_{t \supset e} \frac{\partial a_t}{\partial l_e} \epsilon_t=0  \q .
\ea

\section{Conclusion}\label{conc}

We presented different formulations for Regge calculus with cosmological constant and with simplices of constant sectional curvature. Using simplices with constant curvature leads not only to a better approximation of the continuum but leads often to more elegant formulations for instance in the case of the first order formulation. One particular interesting feature is that one can take as basic variables angles, either the 3d dihedral angles or the 4d dihedral angles. These variables have to be constrained by some gluing conditions. Again curved simplices have the advantage that angles are sufficient to determine the geometry. In this way all the gluing constraints can be expressed by the universal relations between dihedral angles in $n$--simplices and its $(n-1)$--subsimplices, derived in appendix \ref{details}.

These considerations could be useful for the construction of quantum gravity models. The Tuarev-Viro model defines a partition function for 3d gravity with a cosmological constant and its semi--classical limit gives the Regge action for simplices with constant curvature \cite{taylor}. In the quantum model the cosmological constant is accommodated by deforming the $SU(2)$ gauge group of the underlying gauge formulation to the quantum group $SU(2)_{q}$. An open issue is whether a similar construction is possible in 4d, for some steps in this direction see for instance \cite{aristide,lewand}.

To this end a gauge formulation corresponding to the discretized formulations presented here would be useful. One possibility is to consider discretizations of actions using Einstein-Cartan geometries \cite{wise}, as the connection used there leads to the same corrected curvature as for simplices with constant curvature. In particular a formulation similar to the Plebanski action \cite{plebanski} -- corresponding to using 4d and 3d dihedral angles and areas -- could be useful to obtain spin foam quantizations for 4d gravity with cosmological constant. This is not only necessary to match physical reality but could also provide a regularization for the quantum gravity models (i.e. an IR cut--off for positive cosmological constant).



\begin{appendix}

\section{Geometric relations in simplices}\label{appendix}

Consider a $D$--dimensional simplex  in a $D$--dimensional manifold of constant sectional
curvature $\kappa\neq 0$ (i.e. the sphere $S^D$ for positive $\kappa$ and hyperbolic space $H^D$ for negative $\kappa$) consisting of $D+1$ vertices $v_1,\ldots,
v_{D+1}$. Denote this simplex by $(123\ldots D+1)$. Any subsimplex
is determined by the subset $v_{i_1},\ldots,v_{i_n}$ of the vertices
which span this subsimplex, and will therefore be denoted as
$(i_1i_2,\ldots,i_n)$. The subsimplices in curved
space are defined to be the hypersurfaces with zero extrinsic
curvature as embeddings in the geometry of the higher dimensional
simplex. These are in fact also simplices of curvature $\kappa$. An
edge $(ij)$ is then just given by the geodesic connecting $v_i$ and
$v_j$.
Denote the geodesic lengths of the edges $(ij)$ by
$l_{ij}$. Then the $(D+1)\times(D+1)$ matrix $G$ with entries
\begin{eqnarray} \label{gram1}
G_{ij}\;=\;c_\kappa(l_{ij})
\end{eqnarray}
where the function $c_\kappa(x)$ is defined by
\begin{eqnarray*} \label{ckappa}
c_\kappa(x)\;:=\;\left\{\begin{array}{ll}
\cos\big(\sqrt\kappa x\big)&\quad \kappa>0\\[5pt]
\cosh\big(\sqrt{-\kappa} x\big)&\quad
\kappa<0\end{array}\right.
\end{eqnarray*}

\noindent is called the length Gram matrix of the simplex. 
We denote by
$c_{ij}$ the $ij$-th cofactor of $G$, i.e. the determinant of the
matrix obtained by removing the $i$-th row and $j$-th column of $G$ multiplied by $(-1)^{i+j}$.
Then the interior dihedral angle $\theta_{ij}$ opposite the edge
$(ij)$ is given by \cite{kokk}
\begin{eqnarray}\label{dihedral}
\cos\theta_{ij}\;=\;-\frac{c_{ij}}{\sqrt{c_{ii}}\sqrt{c_{jj}}} \q .
\end{eqnarray}
As the cofactor of an invertible matrix is the inverse multiplied by the determinant of the matrix, this formula also holds if we replace the  cofactor by the inverse. The length Gram matrix is invertible for non--degenerate simplices \cite{taylor}.
For a  triangle, say with positive curvature, this reduces to
\ba
\cos\alpha_{ij}=\frac{\cos \sqrt{\kappa} l_{ij} -\cos \sqrt{\kappa} l_{ik}\, \cos\sqrt{\kappa}l_{jk}}{\sin\sqrt{\kappa}l_{ik} \,\sin\sqrt{\kappa} l_{jk}} \q ,
\ea
a relation that we will encounter again between dihedral angles of different dimension.

The angle Gram matrix is defined by $\tilde G_{ij}:=-\cos\theta_{ij}$ for $i\neq j$ and $\tilde G_{ii}=1$. This angle Gram matrix and the length Gram matrix (\ref{gram1}) are in a certain sense dual to each other. Precisely we have \cite{kokk}
\ba
G_{ij}=\frac{\tilde{c}_{ij}}{\sqrt{\tilde c_{ii}}\sqrt{\tilde c_{jj}}}
\ea
where $\tilde c_{ij}$  is the $ij$-th cofactor of $\tilde G$.
In this way we can express the lengths as a function of the dihedral angles. For the triangle with $\kappa>0$ we obtain
\ba
\cos\sqrt{\kappa}l_{ij}=\frac{ \cos\alpha_{ij}+\cos\alpha_{ik}\cos\alpha_{jk}}{\sin\alpha_{ik}\sin\alpha_{jk}} \q .
\ea

 Denote the volume of the
subsimplex spanned by all vertices except $v_i$ and $v_j$ by
$V_{(\hat i\hat j)}$. For variations $\delta$ of the geometry of a simplex the Schl\"afli identity \cite{milnor}
\begin{eqnarray}\label{Gl:SchlaefliIdentityForCurvedSimplices}
\sum_{i<j}V_{(\hat i\hat j)} \delta \theta_{ij}\;=\;(D-1)\,\kappa\,\delta V_{(12\ldots
D+1)}
\end{eqnarray}
 holds.
There is no general explicit formula (not involving integration) for the volume of a simplex with constant curvature available for $D\geq 3$ \cite{luo}. For the variation of the volume term in the action one can however use the Schl\"afli identity (\ref{Gl:SchlaefliIdentityForCurvedSimplices}), that re--expresses this variation as a variation of the dihedral angles.

\subsection{Relations between dihedral angles} \label{details}

We want to relate the dihedral angles $\theta_{ij}$ in a simplex $\sigma$ to the
dihedral angles $\phi_{mn,p}$ in the subsimplex $\sigma(\hat p)$ not
containing the vertex $p$ and opposite the edge $(mn)$. To this end
we consider the Gram matrix $H_{ij}(p)$ of the subsimplex
$\sigma(\hat p)$ whose entries coincide with those of $G_{ij}$ for
$i,j \neq p$. It is straightforward to express the inverse $H^{kl}(p)$ of
$H_{ij}(p)$ using the inverse $G^{kl}$ of $G_{ij}$:
\eqa\label{ba1}
H^{kl}(p)=G^{kl}-\frac{G^{pk} G^{lp}}{G^{pp}}  \q .
 \neqa
 Inserting
the definition of dihedral angles  (\ref{dihedral}) we obtain
\eqa\label{ba2} \cos\phi_{lm,p}=
\frac{\sqrt{G^{ll}G^{mm}}}{\sqrt{H^{ll}
H^{mm}}}\left(\cos\theta_{lm}+\cos\theta_{pl}\cos\theta_{pm}\right).
  \,\,\,\neqa
From (\ref{ba1}) we calculate \ba H^{ll}(p)=G^{ll}-\frac{G^{pl}
G^{lp}}{G^{pp}} = G^{ll} \,(1- \cos^2 \theta_{lp} )\q. \ea Hence
(\ref{ba2}) turns into
\ba\label{bfinal} \cos\phi_{lm,p}&=&
\frac{\cos\theta_{lm}+\cos\theta_{pl}\cos\theta_{pm}}{\sin\theta_{pl}\sin\theta_{pm}}\,
. \ea
The same relation holds for flat simplices \cite{bdss}. The
inversion of formulas (\ref{bfinal}) yields
\ba
\cos\theta_{lm}&=&
\frac{\cos\phi_{lm,p}-\cos\phi_{lp,m}\cos\phi_{mp,l}}{\sin\phi_{lp,m}\sin\phi_{lm,p}}
\, . \ea
This allows to express the 4d deficit angles as a
function of the 3d dihedral angles (if one specifies the subsimplex $\sigma(\hat p)$ for every dihedral angle $\theta_{lm}$).

Note that we derived the relations (\ref{bfinal}) for simplices of
arbitrary dimension. Hence it holds also between the 3d dihedral
angles $\phi_{lm,p}$ in a tetrahedron $\sigma(\hat p)$ and the 2d
dihedral angles $\alpha_{lm,kp}$ in the triangles $\sigma(\hat k \hat p)$ in a 4-simplex $\sigma$:
\eqa\label{ba3}
\cos\alpha_{lm,kp}=\frac{\cos\phi_{lm,p}+\cos\phi_{kl,p}\cos\phi_{km,p}}{\sin\phi_{kl,p}\sin\phi_{km,p}}
\q . \neqa

To calculate the areas as a function of the dihedral angles
$\phi$ we can invoke the formula
\ba
\kappa\, a_t= \alpha_{12}+ \alpha_{23} +\alpha_{31} -\pi,
\ea
 that expresses the area
as a function of the three 2d dihedral angles $\alpha_{ij}$, which in turn
can be expressed as a function of the 3d angles $\phi$ with
(\ref{ba3}). Again these $\phi$ can be expressed as functions of the 4d dihedral angles $\theta$, see (\ref{bfinal}), so that one can express $a_t$ as a function of the 3d angles in a tetrahedron or the 4d angles in a 4--simplex.
Similarly the volume of a $4$-simplex can (in principle) either be
expressed as a function of the ten 4d angles $\theta$ or the ten lengths
of the simplex. The 4d angles and the lengths can in turn be
expressed as functions of the $\phi$ and consistency of this
procedure is guaranteed by the constraints (\ref{ba4}).

\end{appendix}

\subsubsection*{Acknowledgements}
We thank
John Barrett, Valentin Bonzom, Daniele Oriti and Simone Speziale for discussions and Freddy Cachazo for inspiring the title.

\appendix



\end{document}